# Correlation between site preference of ternary Mn addition in LaAg and superconductivity

**S. Kumar and S. N. Kaul**[*]
*School of Physics, University of Hyderabad, Hyderabad – 500 046, India*

**J. Rodríguez Fernández** and **L. Fernández Barquín**
*CITIMAC, Universidad de Cantabria – 39005 Santander, Spain*

**P. F. Henry**
*Institute Laue Langevin, BP 156, 38042 Grenoble, France*

The results of an extensive investigation of structure, surface morphology, composition and the superconducting-normal phase diagram of a new unconventional superconductor $LaAg_{1-c}Mn_c$ with nominal composition c = 0.0, 0.025, 0.05, 0.1, 0.2 and 0.3, reveal the following. The alloys with $c$ = 0, 0.025 and 0.05 are essentially single phase alloys with the actual Mn concentration, $x$, same as the nominal one, i.e., $c = x$, whereas in the alloys with $c$ = 0.1, 0.2 and 0.3, the actual Mn concentration of the majority phase (crystalline grains) is $x$ = 0.050(1), 0.080(1) and 0.100(1), respectively. The ternary Mn addition does not alter the CsCl structure of the parent compound LaAg. Neither a structural phase transition occurs nor a long-range antiferromagnetic order exists at any temperature within the range $1.8\,K \leq T \leq 50\,K$ in any of the Mn containing alloys. Mn has exclusive La (Ag) site preference in the alloy (alloys) with $x = c = 0.025$ ($x \geq 0.05$ or c $\geq$ 0.1) whereas in the alloy with $x = c = 0.05$, Mn has essentially no site preference in that all the Mn atoms either occupy the La sites or the Ag sites. In the alloys (alloy) with $x \geq 0.05$ ($x = c = 0.025$), substitution of Ag (La) by Mn at the Ag (La) sub-lattice sites in LaAg host gives rise to unconventional superconductivity (destroys the conventional phonon-mediated superconductivity prevalent in the parent LaAg compound).
PACS numbers: 61.12.Ld; 74.70.Dd; 81.05.Bx; 74.25.Dw

[*]*E-mail*: kaul_sn@yahoo.com

# I. INTRODUCTION

Unconventional superconductivity has been recently reported[1] in a new nearly antiferromagnetic metal[2], i.e., in the ternary LaAg$_{1-c}$Mn$_c$ alloys. Compared to the heavy-Fermion systems, which are known[3,4] to exhibit this form of superconductivity, the superconducting transition temperature, $T_C$, is nearly 10 times higher and superconductivity is robust against impurities in these alloys. Moreover, the ternary Mn addition in the parent LaAg compound (which crystallizes into the CsCl structure with La and Ag atoms respectively occupying the body-centred and corner sites) increases the superconducting transition temperature abruptly from $T_C \approx 1K$ (in[5] LaAg) to $\cong 5K$ when $c \geq 0.05$. However, the actual mechanism responsible for an abrupt increase in $T_C$ when the Mn concentration in La-Ag-Mn alloys exceeds the threshold value of 5 at. % is not clear at present. In order to gain physical insight into the mechanism of electron pairing in the alloys in question, we recall a few known facts.

When Ag is added to La in the equiatomic proportion to form the intermetallic compound LaAg, $T_C$ of the double-hexagonal-close-packed phase of La drops from[6] $\approx$ 5K to[5] $T_C \approx$ 1K. LaAg compound is a Pauli spin paramagnet[7] in the normal state and has a low $T_C$ because the CsCl structure is one of the most unfavorable structures[8] for the occurrence of conventional phonon-mediated superconductivity. Conventional phonon-mediated superconductivity in La element is known to be extremely sensitivity to even a small concentration of *magnetic* impurities, i.e., a concentration of magnetic impurities as small as $1\,at.\%$ suffices to suppress superconductivity[9] in La completely. Based on this observation, the superconductivity prevalent in the LaAg host, is expected to get completely destroyed if the concentration of magnetic Mn atoms (each Mn atom carries a magnetic moment of[2] 4 μ$_B$) substituting La on the La sub-lattice sites in LaAg exceeds 1 at %. These considerations strongly suggest that the superconductivity observed in LaAg$_{1-c}$Mn$_c$ alloys with $c \geq 0.05$ could be a consequence of the substitution of Ag by Mn at the Ag sites unless Mn substitution alters the CsCl structure of the parent compound LaAg. The knowledge of whether or not the CsCl structure persists to temperatures T « T$_C$ in the Mn containing alloys and whether the *magnetic* Mn atoms preferentially occupy the La or Ag sites or have no preference at all for any site, is thus crucial to understanding the nature of superconductivity in La-Ag-Mn alloys. This realization prompted us to undertake a detailed investigation of structure, surface morphology and composition in $LaAg_{1-c}Mn_c$ alloys.

## II. EXPERIMANTAL DETAILS

Recognizing that superconductivity is extremely sensitive to the local composition fluctuations, imperfections and impurities (and hence to the sample preparation conditions), the $LaAg_{1-c}Mn_c$ alloy samples of nominal Mn composition $c = 0.00, 0.025, 0.05, 0.1, 0.2$ and $0.3$ were prepared by two different techniques: induction-melting and arc-melting. High-purity La (99.98%), Ag (99.999%) and Mn (99.99%) in the form of cylindrical pieces, wire pieces and platelets, taken in stoichiometric proportions by weight, were inserted into a tungsten tube of 1 cm inner diameter and 10 cm length closed at one end, installed in an arc furnace. The open top end of the tungsten tube was sealed with a tungsten cap (by striking an arc between tungsten electrode and the rim of the tungsten cap) under a positive pressure of 99.999% pure argon gas after evacuating and flushing the tungsten tube several times with 5N pure argon gas. The sealed tungsten tube was removed from the arc furnace and installed in a graphite susceptor, placed within the induction coils of the radio-frequency (RF) induction-melting set-up. The contents of the sealed tungsten tube were melted at $1400^oC$ using RF induction heating. The molten alloy (the melt in the tube) was kept at $1400^oC$ for nearly an hour for homogenization and then furnace-cooled by turning off the RF power. The tungsten tube was broken open and the sample, in the form of a shinning cylindrical rod (of diameter 1 cm and length 3 cm), was retrieved with ease. No reaction of the contents with the inner wall of the tungsten tube was detected. Polycrystalline samples, so prepared, are henceforth referred to as the IM (induction-melted) samples. The samples of nominal composition $c = 0.1, 0.2$ and $0.3$ were prepared by this technique. In order to prepare the samples of nominal Mn composition $c = 0.00, 0.025, 0.05, 0.1, 0.2$ and $0.3$ by the arc-melting technique, the high-purity elements La, Ag and Mn, taken in appropriate proportions by weight, were placed on a water-cooled copper hearth in an arc furnace and arc-melted under 99.999% pure argon gas atmosphere. The alloy buttons, so formed, were repeatedly turned upside down and re-melted with a view to improve their compositional homogeneity. The alloy samples prepared by this method are henceforth labeled as the AM (arc-melted) samples. The IM rods and AM buttons were spark-cut into samples of desired sizes and shapes for different measurements. Prior to a given measurement, sample surfaces were mechanically polished to remove the surface oxidation layer and then cleaned with acetone.

X-ray diffraction (XRD) patterns for all the compositions were recorded at room temperature in the θ - 2θ scattering geometry at 0.02 Å steps in the 2θ range from 10° to 100° on a Philips powder diffractometer using monochromatic Cu $K_\alpha$ radiation. Unlike x-rays, neutrons penetrate the entire specimen and as such neutron diffraction (ND) experiments yield the bulk structural information. High-resolution ND measurements were performed at 2 K intervals in the temperature range $1.8\,K \leq T \leq 50\,K$ at the neutron wavelengths of $\lambda$ = 1.297 Å and 2.408 Å and steps of $0.1^o$ in the scattering angle range of $1^o$ to $157.6^o$ on thermal neutron high flux two-axes D20 diffractometer at the Institut Laue-Langevin in Grenoble. High-flux D20 neutron diffractometer enables the detection of even the most feeble magnetic and/or nuclear reflections and thereby permits an accurate determination of even the subtle changes in the magnetic order and atomic structure in the samples under study. The samples of cylindrical shape were inserted into thin-walled cylindrical vanadium cans of diameter 6 – 8 mm (depending on the sample diameter) and height 6 cm, mounted inside a bath-type liquid helium cryostat. The sample temperature was controlled by a proportional-Integration-differentiation (PID) temperature controller and was stable to within 0.1 K during the measurement period. The temperature range covered in the neutron diffraction experiments embraces the superconducting transition temperatures for the samples with nominal Mn concentration $c \geq 0.05\,at.\%$.

The Rietveld method was used to refine the crystal structure. The XRD and ND data were analyzed using the FULLPROF profile refinement program[9] which permits the simultaneous determination of several crystallographic and magnetic coexisting phases. A pseudo-Voigt function was used to describe the line profile.

## III. RESULTS AND DISCUSSION

### A. Scanning electron microscopy and energy dispersive absorption of x-rays

Scanning electron microscopy (SEM) has been used to ascertain the number of crystalline phases present in the La-Ag-Mn alloy samples and the energy dispersive absorption of x-rays (EDAX) analysis for determining the composition of the phases. The scanning electron micrographs of all the samples, displayed in the figures 1 and 2, clearly highlight vast, relatively bright, regions or patches (crystalline grains, the so-called *majority phase*) surrounded by dark lines (the grain boundaries); or spots or small black regions (the *minority phases*). The results (the average of at least 6 readings) of a detailed compositional evaluation of bright patches, small black patches (nearly rectangular in shape) and dark boundary lines or spots, marked as regions I, II and III, respectively, in the SEM pictures (figures 1 and 2), by EDAX, are displayed in Table I and summarized as follows. (i) The global average composition (i.e., the composition recorded when the EDAX window encompasses all the different regions, I and II or I – III, in the SEM picture of a given sample) tallies with the nominal composition for all the alloy compositions studied. (ii) In the SE micrograph (Fig.1(a)) of the parent compound LaAg, except for a few small dark rectangular regions II, which are *rich* in La (~ 95%), the entire remaining region has a *uniform composition* of $La_{1.00(1)}Ag_{1.00(1)}$. (iii) Similarly, in the micrograph (Fig. 1(b)) of the alloy with c = 0.025, the composition of the bright regions I is $La_{1.000(1)}Ag_{0.975(3)}Mn_{0.025(2)}$ and that of the dark regions II is $La_{1.010(1)}Ag_{0.976(2)}Mn_{0.024(1)}$, the sample is thus a *single phase alloy* with the composition $La_{1.000(1)}Ag_{0.975(3)}Mn_{0.025(2)}$. (iv) The composition of the regions I (bright) and II (dark) in the scanning electron micrograph, Fig. 1(c), of the 'as-prepared' sample with c = 0.05 is $La_{0.950(1)}Ag_{1.00(1)}Mn_{0.050(2)}$ and $La_{1.00(1)}Ag_{0.950(1)}Mn_{0.050(2)}$, respectively. By contrast, the EDAX analysis of the white regions I (grains) and dark lines or spots (grain boundary regions II) in the SEM picture, Fig.1(d), of the c = 0.05 alloy sample, annealed at 600$^o$C for 7 days, reveals that the *crystalline grains* in the *annealed sample* have the *same* Mn content as that in the as-prepared sample with c = 0.025 (Table I), and that the remaining amount of Mn (nearly half of the original Mn content in the as-prepared sample) has segregated out to the grain boundaries, where, as a consequence, a Mn-rich phase forms.

Table I. Nominal and actual chemical composition of the samples. a – 'arc-melted'; b – 'as-cast'; c – 'annealed'; d – 'induction-melted'

| Nominal Composition LaAg$_{1-c}$Mn$_c$ | | | Actual Composition | | | | | | | | |
|---|---|---|---|---|---|---|---|---|---|---|---|
| | | | Majority Phase Region I | | | Minority Phase Region II | | | Minority Phase Region III | | |
| La | Ag | Mn | La | Ag | Mn | La | Ag | Mn | La | Ag | Mn |
| 1.00 | 1.00 | 0.00[a, b] | 1.00(1) | 1.00(1) | 0.00 | 1.920(5) | 0.080(5) | 0.00 | - | - | - |
| 1.00 | 0.975 | 0.025[a, b] | 1.000(5) | 0.975(5) | 0.025(1) | 1.010(5) | 0.976(5) | 0.024(1) | - | - | - |
| 1.00 | 0.95 | 0.05[a, b] | 0.950(1) | 1.00(1) | 0.050(2) | 1.000(1) | 0.950(1) | 0.050(2) | - | - | - |
| 1.00 | 0.95 | 0.05[a, c] | 0.975(3) | 1.00(1) | 0.025(2) | - | - | - | 0.900(1) | 0.900(1) | 0.200(1) |
| 1.00 | 0.90 | 0.10[a, b] | 0.95(1) | 1.00(1) | 0.050(2) | 1.950(1) | 0.030(1) | 0.020(1) | 0.800(1) | 0.800(1) | 0.400(1) |
| 1.00 | 0.90 | 0.10[d, b] | 0.950(5) | 1.000(5) | 0.050(1) | 0.091(1) | 0.035(1) | 1.875(1) | 0.800(1) | 0.800(1) | 0.400(1) |
| 1.00 | 0.80 | 0.20[a, b] | 0.92(1) | 1.000(1) | 0.080(2) | 1.500(1) | 0.450(1) | 0.050(1) | 0.900(1) | 0.900(1) | 0.200(1) |
| 1.00 | 0.80 | 0.20[d, b] | 0.920(5) | 1.000(5) | 0.080(1) | 1.800(1) | 0.030(1) | 0.170(1) | 0.750(1) | 0.750(1) | 0.500(1) |
| 1.00 | 0.70 | 0.30[d, b] | 0.900(5) | 1.000(5) | 0.100(1) | 1.800(1) | 0.170(1) | 0.030(1) | 0.200(1) | 0.200(1) | 1.600(1) |

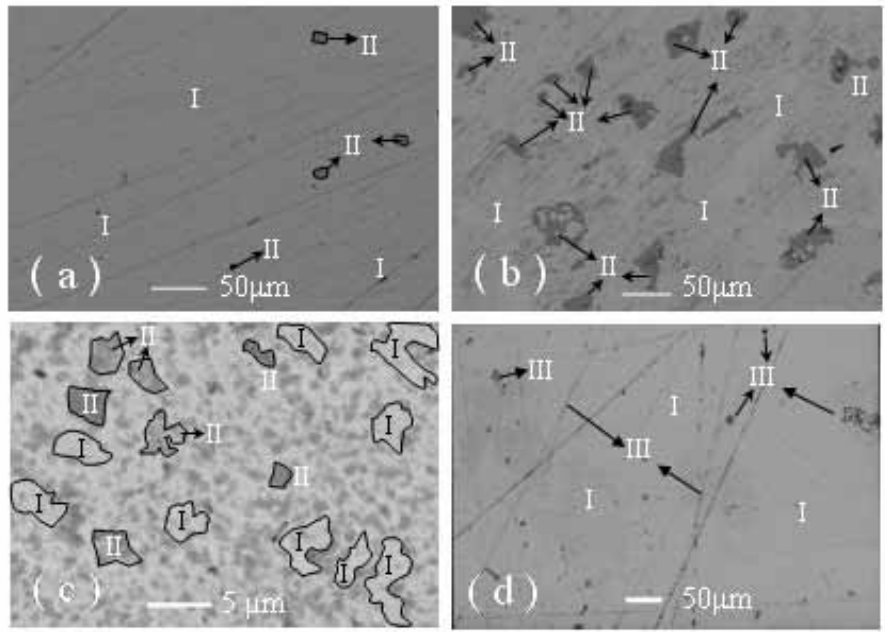

FIG.1. Scanning electron micrographs for the alloys (a) c = 0.0, (b) c = 0.025, (c) c = 0.05 (as-cast) and (d) c = 0.05 (annealed).

(v) Fig.2(a) displays the micrograph for the AM samples with c = 0.1 while Fig.2(b) corresponds to the IM counterpart. In both AM and IM samples, the crystalline grains (bright regions I) are separated by the dark lines or dotted patches (regions III), which are the grain boundaries, and occasionally contain square or triangular dark patches (regions II). EDAX analysis of the regions I – III reveals that the grains (the majority phase) have only half of the nominal Mn content, i.e., their actual composition is $La_{0.95(1)}Ag_{1.00(1)}Mn_{0.050(2)}$ and, like in the *annealed* c = 0.05 sample, the remaining Mn segregates out to grain boundaries and to regions II where it forms a Mn-rich phase of composition $La_{1-x'}Ag_{1-x'}Mn_{2x'}$ with $x'=0.2$, i.e., $La_{0.80(1)}Ag_{0.80(1)}Mn_{0.40(1)}$. (vi) In the micrographs for the alloys with $c$ = 0.2 (AM) and $c$ = 0.3 (IM), shown in Fig. 2(c) and 2(d) respectively, the composition of the crystalline grains (region I) is $La_{0.92(1)}Ag_{1.00(1)}Mn_{0.080(2)}$ and $La_{0.900(5)}Ag_{1.000(5)}Mn_{0.100(1)}$ and that of the grain boundaries (dark lines or spotted patches; region III) is $La_{1-x'}Ag_{1-x'}Mn_{2x'}$ with $x'=0.1$ and 0.8 for $c$ = 0.2 and $c$ = 0.3, respectively.

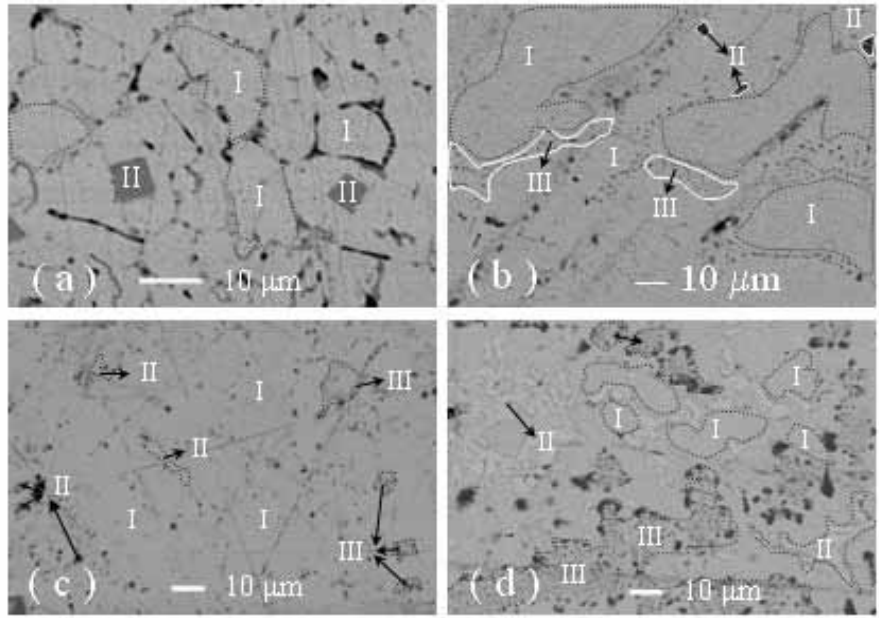

FIG.2. Scanning electron micrographs for the alloys with (a) c = 0.1 (AM), (b) c = 0.1 (IM), (c) c = 0.2 (AM) and (d) c = 0.3 (IM).

Thus the EDAX analysis permits us to conclude that the alloys with $c$ = 0, 0.025 and 0.05 are *essentially single phase* alloys with the actual Mn concentration, $x$, matching the nominal one, i.e., $c = x$, whereas in the alloys with $c$ = 0.1, 0.2 and 0.3, the actual Mn concentration of the majority phase (crystalline grains) is $x$ = 0.050(1), 0.080(1) and 0.100(1), respectively.

## B. X - ray diffraction

Figure 3(a) compares the x-ray diffraction (XRD) patterns taken at room temperature on disc-shaped samples of the $LaAg_{1-c}Mn_c$ alloys with c = 0.00, 0.05, 0.1, 0.2 and 0.3. The results of a preliminary analysis of these patterns were reported earlier[1]. However, a detailed analysis of these diffraction patterns was carried out by the Rietveld method, using the FULLPROF profile refinement program, in the present work. All the observed Bragg peaks, including the weakest reflections, in a given XRD pattern could be completely indexed on the basis of cubic CsCl (space group *Pm-3m*) structure of the parent LaAg compound. That this is indeed the case is clearly brought out by a comparison between the Rietveld-refined (continuous lines) and observed (filled circles) XRD patterns for the alloys with c = 0.05 and c = 0.2 in figure 3(b). In this figure, the vertical bars below the XRD patterns mark the positions of the calculated Bragg peaks while the difference between the observed and calculated patterns, $\Delta I = I_{obs} - I_{calc}$, is shown in the lower panel. Such an agreement between the calculated and observed XRD profiles is representative of all the samples. The salient features of the XRD patterns, shown in Fig. 3, as revealed by the Rietveld analysis, are as follows. (a) CsCl structure of the parent compound LaAg accounts for all the Bragg peaks observed in all the Mn containing alloys. This observation places an upper bound of 5 volume percent of the main CsCl phase on any other minority phase (s), if present. (b) The diffraction patterns of the samples $x = c = 0.05$ and $c = 0.1$ exactly coincide, with the result that the lattice parameter has the same value (within the uncertainty limits) for these two alloys. This observation thus confirms the previous EDAX result that the majority phase in the alloy with $c = 0.1$ has actual composition $x = 0.05$. (c) After peaking at $x = c = 0.025$, the lattice parameter '$a$' has a *linear* variation with the *actual* Mn concentration, $x$, in the range $0 \leq x \leq 0.1$ except for the value at $x = c = 0.025$. This is evident from figure 4, which displays the variation of the lattice parameter '$a$' with the actual Mn concentration, as determined from the Rietveld refinement of the observed x-ray diffraction patterns.

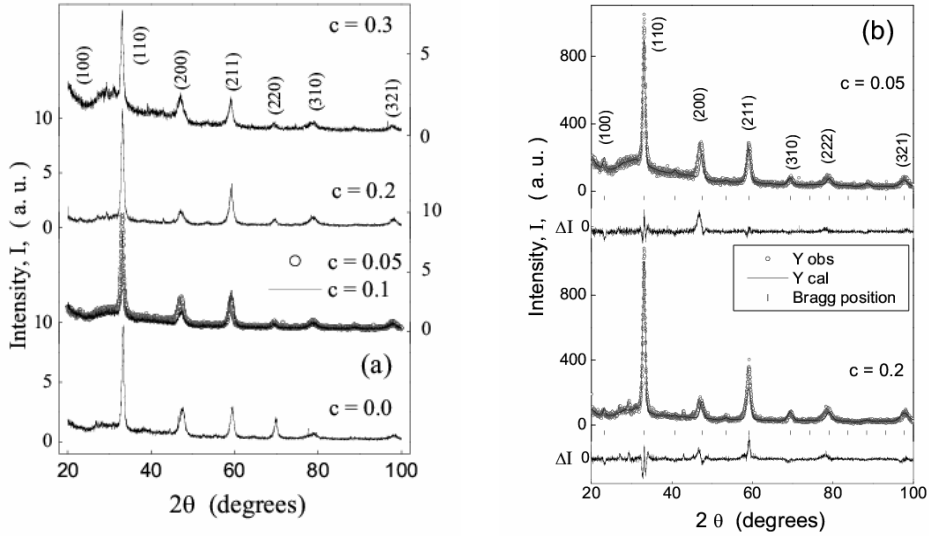

FIG.3. (a) X-ray diffraction patterns taken at room temperature on the alloys with $0 \leq c \leq 0.3$.
(b) Typical Rietveld refinement of the XRD patterns.

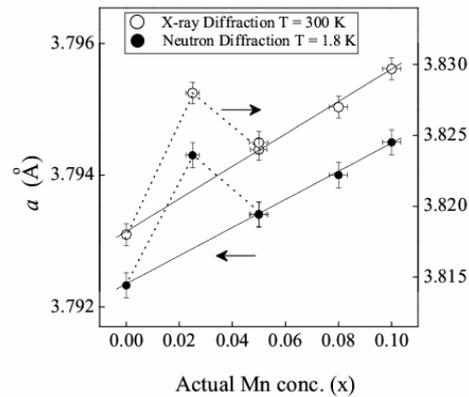

FIG.4. Lattice parameter 'a' as a function of actual Mn concentration 'x'.

## C. Neutron diffraction

Figures 5 – 8 show typical neutron diffraction (ND) patterns over the scattering angle range 15° to 145° observed at neutron wavelengths $\lambda = 1.297$ Å and $\lambda = 2.408$ Å and temperatures 1.8 K and 50 K along with the best least-squares Rietveld fits (continuous lines). A normalized sum of the least squares, $\chi^2$, and the weighted refinement factor, $R_{wp}$, defined as

$$\chi^2 = \frac{\sum_i ((I_o - I_c)/\sigma I_o)^2}{N_o - N_p} \qquad (1)$$

and

$$R_{wp} = 100 \times \left\{ \frac{\sum_i w_i (I_o - I_c)^2}{\sum_i w_i I_o^2} \right\}^{1/2} \qquad (2)$$

were used to measure the quality of the fits obtained in the profile refinements. In equations (1) and (2), summation is over $N_o$ measured data points, $I_o$ and $I_c$ are the observed and calculated intensities, $\sigma I_o$ is the estimated standard deviation of $I_o$, $N_p$ is the number of refined (free fitting) parameters and $w_i = 1/I_o^i$ are the weight factors. The cell parameter, intensity, the half-widths U, V, W, occupancies, thermal parameters, strain parameter and the zero point of the diffractometer (i.e., a total of 14 parameters) were refined for the pattern recorded at a given temperature keeping the asymmetry parameters fixed at the values obtained at the highest temperature. This Rietveld refining procedure not only yielded excellent fits to all the diffraction patterns with typical values for the goodness of fit index, $S = R_{wp}/R_e \leq 2$ (where $R_e$ is the expected refinement factor), $2 \leq \chi^2 \leq 4$ and $25\% \leq R_{wp} \leq 30\%$ but also resulted in the variation in the occupancy of Mn atoms on the La and Ag sites with the Mn concentration that is *consistent* with the EDAX results. In Figs. 5, 7, 8, the vertical bars in the first row, below a given diffraction pattern, mark the positions of the calculated Bragg peaks corresponding to the crystalline phase with CsCl structure of the parent compound LaAg whereas those in the second row correspond to the body-centred cubic phase of Vanadium (these extremely weak Bragg reflections, present in the neutron diffraction patterns of all the samples, originate from the vanadium cans containing these samples). The difference between the observed and calculated patterns, $\Delta I = I_o - I_c$, shown in the lower panel, assert that all the observed Bragg peaks (except for two very tiny peaks of the vanadium cans) can be indexed based on the CsCl structure. These difference plots, therefore, confirm that the samples are essentially single phase alloys. The slight discrepancy between the observed and calculated patterns can be traced back to the presence of other minority crystalline phases or the partial crystallographic texture in the 'as-cast' samples. Note that the *powder* diffraction patterns could not be recorded since the malleable nature of the samples prevented us from grinding the 'as-cast' samples into powders.

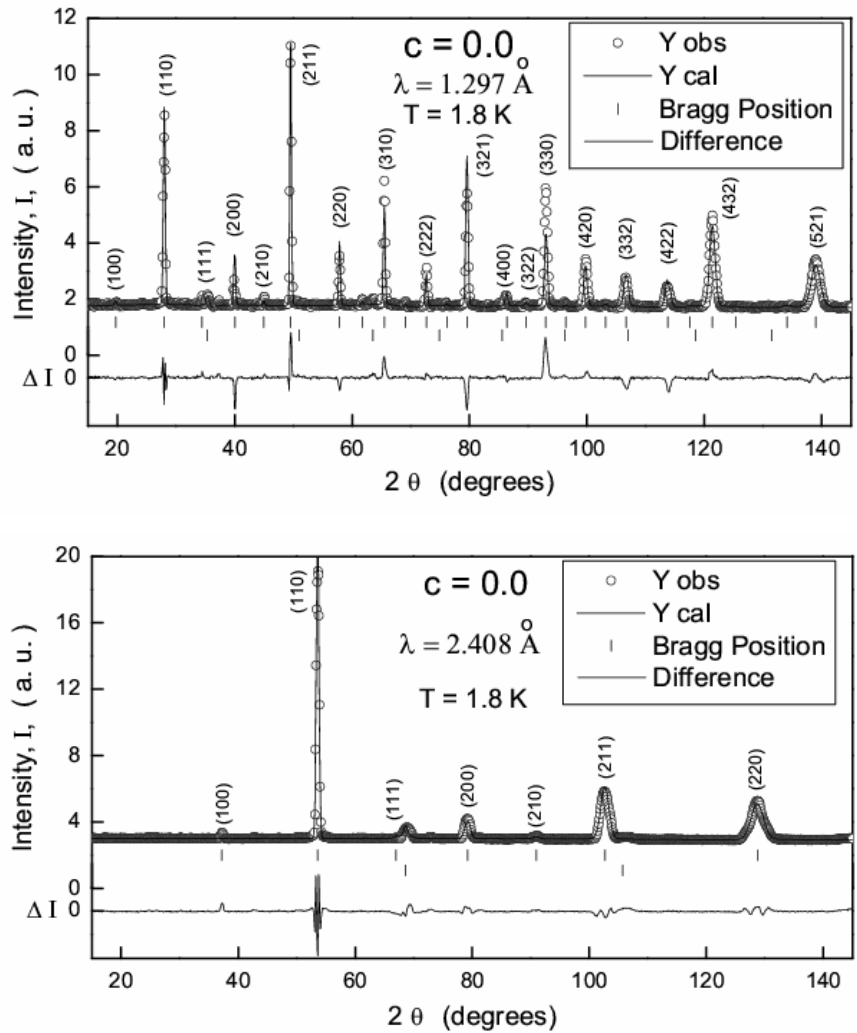

FIG.5. Observed (open circles) and Rietveld refined (continuous lines) neutron diffraction patterns for LaAg at T = 1.8 K and neutron wavelengths λ = 1.297 Å and λ = 2.408 Å.

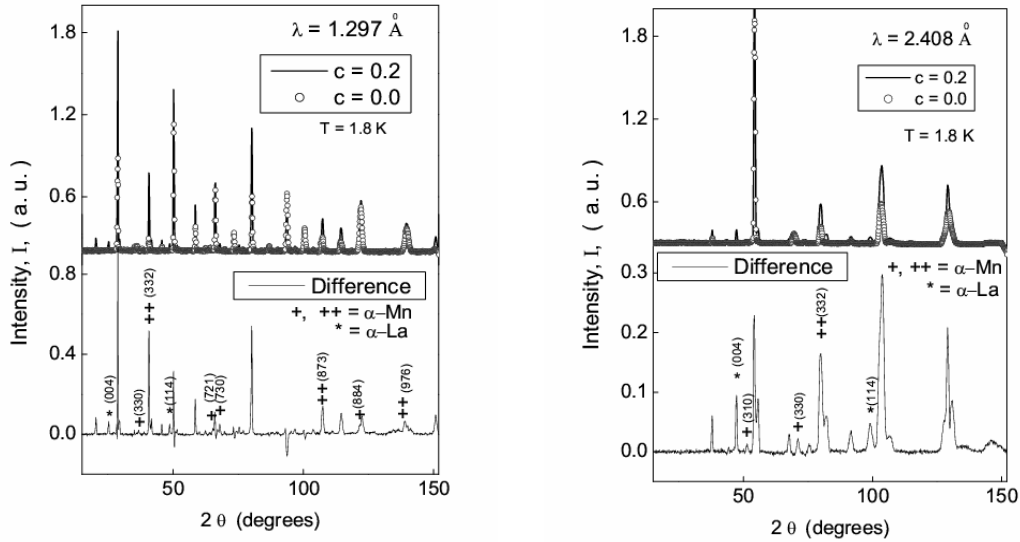

FIG.6. Neutron diffraction patterns of LaAg (c = 0) and the alloy with c = 0.2 taken at T = 1.8 K and neutron wavelengths λ = 1.297 Å and 2.408 Å. The difference pattern = ND pattern of c = 0.2 − ND pattern of c = 0, is shown in the lower panel. The extra peaks in the difference ND pattern are indexed as shown.

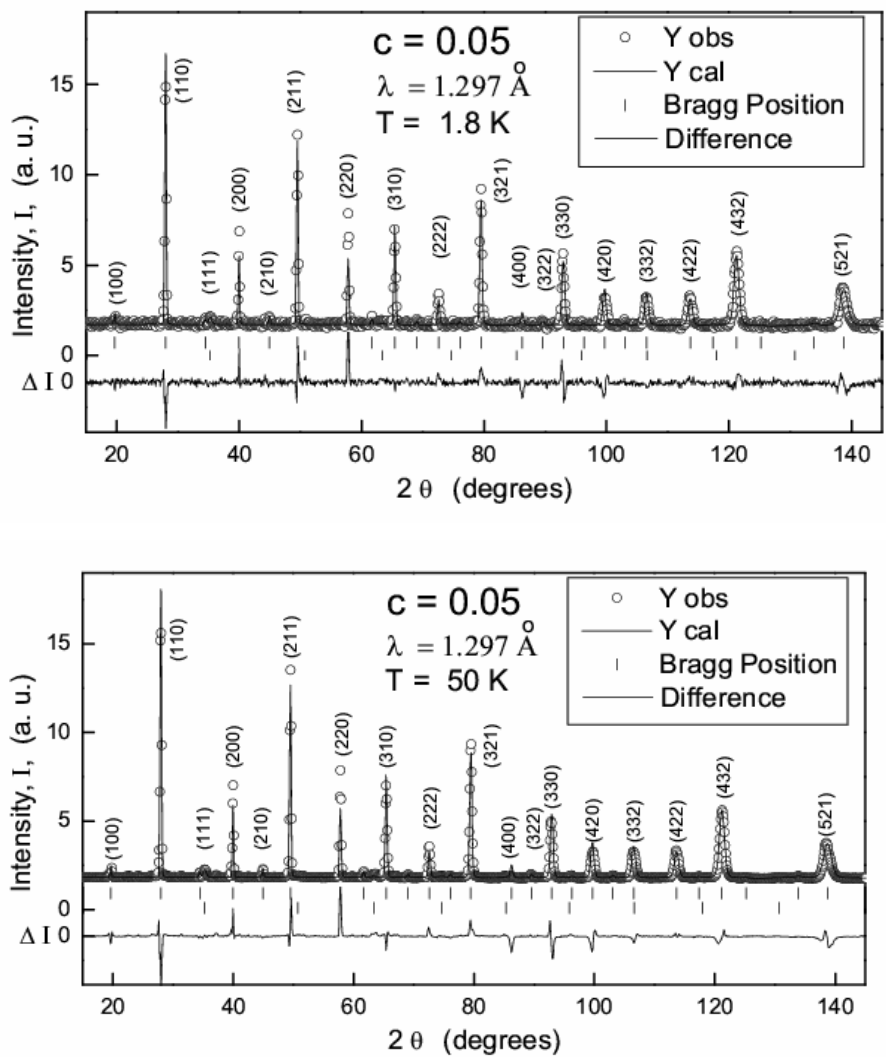

FIG.7. Observed (open circles) and Rietveld refined (continuous lines) neutron diffraction patterns for the alloy with c = 0.05 at neutron wavelength λ = 1.297 Å and temperatures 1.8 K and 50 K.

In order to ascertain whether or not other minority phases are present, a detailed phase analysis, using the Rietveld refining, was undertaken. Making use of the information about the composition of various minority phases obtained from the EDAX analysis (Table I), the Rietveld refinement of the observed ND patterns was carried out by including all possible crystalline phases of La, Ag and Mn, besides the CsCl LaAg and cubic vanadium phases, particularly in the samples with high Mn content, i.e., $c \geq 0.1$, in which a substantial amount of Mn segregates out to grain boundaries and to other regions creating, in the process, La- and Mn-rich regions. This exercise revealed that some of the extremely weak Bragg peaks, which can barely be distinguished from the background, could be accounted for based on the hexagonal $\alpha-La$ (space group: $P6_3/mmc$) and body-centred cubic $\alpha-Mn$ (space group: $I4/mmm$) phases and that all such minority phases put together amount to less than one volume percent. Thus, the alloys with $c \geq 0.1$ are $\approx 99\%$ single (majority) phase alloys.

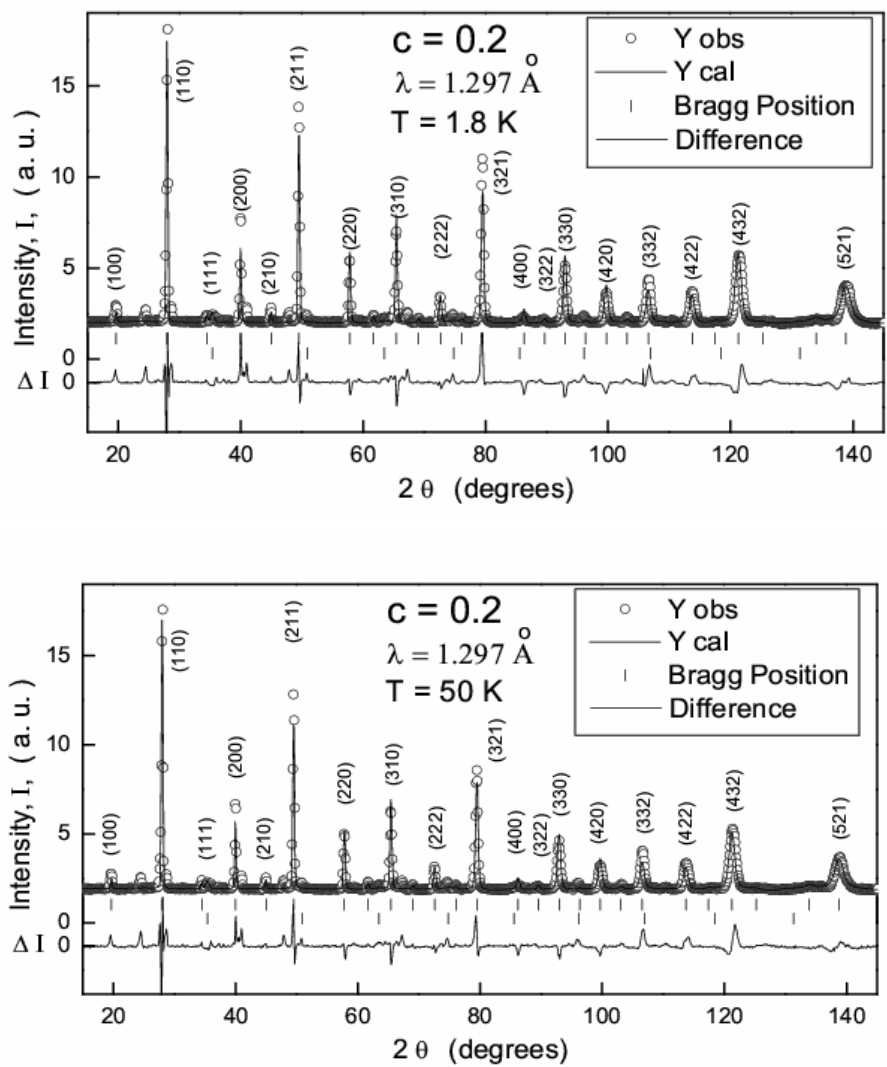

FIG.8. Observed (open circles) and Rietveld refined (continuous lines) neutron diffraction patterns for the alloy with c = 0.2 at neutron wavelength λ = 1.297 Å and temperatures 1.8 K and 50 K.

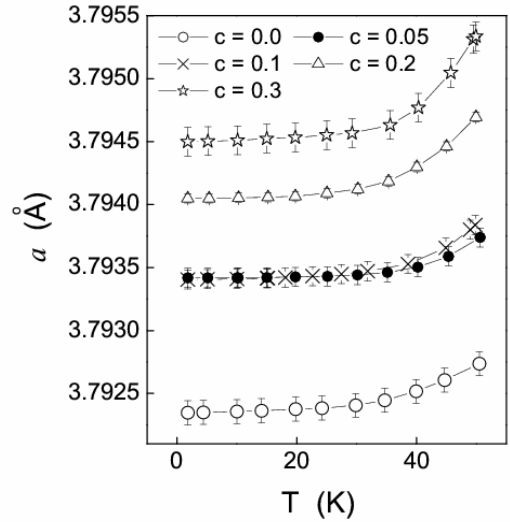

FIG.9. Lattice parameter 'a' as function of temperature for different alloys.

A comparison between the ND patterns of LaAg and the alloy with c = 0.2 (Fig. 6) taken at T = 1.8 K and the neutron wavelengths λ = 1.297 Å and 2.408 Å, representative of all other Mn containing alloys as well, highlights that no magnetic superstructure peaks, symptomatic of long-range antiferromagnetic order, occur at a lower value of momentum transfer, $q = (4\pi/\lambda)\sin\theta$, or equivalently, at low angles at a given value of $\lambda$ even in the alloy with the highest Mn content. The difference plots, shown in the lower panel of figure 6, also reveal the presence of extremely weak Bragg peaks arising *exclusively* from α - Mn (marked by plus sign) and α - La (marked by astrick) phases. On the other hand, α - Mn phase makes a very *minor* contribution to the Bragg peaks marked by ++ which mainly arise from the CsCl phase. Similarly, the ND patterns displayed in Figs. 7 and 8 (taken on the alloys with $x = c = 0.05$ and $c = 0.2$ at the neutron wavelength λ = 1.297 Å and the end temperatures 1.8 K and 50 K of the temperature range covered in such experiments) serve to demonstrate that no extra Bragg peaks (particularly at low scattering angles), indicative of a structural phase transition to a phase of low crystallographic symmetry, appear for any sample at any given $\lambda$, as the temperature is lowered to 1.8 K. The Rietveld profile refinement of the ND patterns taken at a fixed temperature in the range $1.8\,K \leq T \leq 50\,K$ with the neutron wavelengths λ = 1.297 Å and 2.408 Å yields *identical* values of the unit cell parameter '*a*' for a given composition. Figure 4 compares the composition dependence of the cell parameter at 1.8 K, deduced from the ND patterns using the above refining procedure, with that obtained at 300 K by the same method from the XRD patterns. It is evident that the variation of '*a*' with composition yielded by the XRD data is reproduced in facsimile by the ND experiments albeit at a lower temperature. The temperature variations of the lattice parameter, *a*(T), for various alloy compositions, presented in figure 9, clearly demonstrate that '*a*' varies smoothly with temperature and does not exhibit any *discontinuity*, *symptomatic* of a structural phase transition, at any temperature (including the superconducting transition temperature[1], $T_C \cong 5K$) within the temperature range $1.8K \leq T \leq 50K$ covered in the present ND experiments. This observation rules out completely the possibility of a structural phase transition occurring at any temperature in the range 1.8K ≤ T ≤ 50K. That indeed no structural phase transition occurs is further confirmed by the typical difference plots shown in figure 10. In these plots, the observed neutron diffraction pattern recorded at λ = 1.297 Å and T = 50 K is subtracted from that taken at the same wavelength and T = 1.8 K and the difference neutron scattering intensity is plotted against the scattering angle, $2\theta$. It is immediately noticed that the number and positions of the Bragg peaks remains the same between 1.8 K (i.e., in the superconducting state[1], $T < T_C \cong 5K$) and 50 K (i.e., in the normal state[1] where the samples exhibit paramagnetism[2]). The result that no extra magnetic (structural) peaks are observed at low q (low scattering angles) at low temperatures completely rules out the existence (occurrence) of a long-range antiferromagnetic order (structural phase transition).

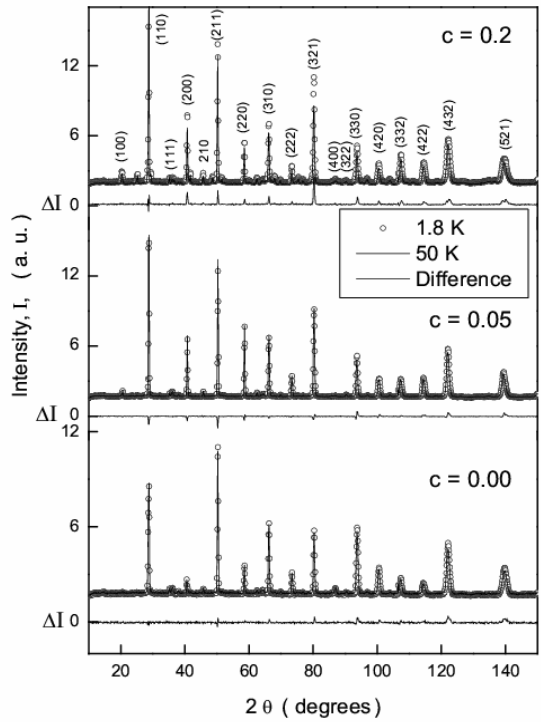

FIG.10 Neutron diffraction plots taken at λ = 1.297 Å and temperatures T = 1.8 K and 50 K. The difference, $\Delta I = I_{1.8K} - I_{50K}$, plots are shown in the lower panel.

The existence of superstructure Bragg reflection peaks in the diffraction pattern, besides the fundamental peaks, basically reflects the presence of long-range atomic order in the sample. To elucidate this point further, we take the example of the CsCl structure of LaAg compound. In this compound, the fundamental Bragg peaks arise from the underlying b.c.c. structure irrespective of the identity of the atoms occupying the body-centred and corner lattice sites of the cubic unit cell. By contrast, the superstructure Bragg peaks appear only when the body-centred and corner sites are preferentially occupied by distinctly different species of atoms, e.g., La (Ag) atoms occupying body-centred (corner) sites and hence their intensities relative to those of fundamental Bragg peaks are a direct measure of atomic order in the system. The long-range order parameter S has been estimated from the observed *integrated* intensities $I_S$ and $I_F$ of the superstructure and fundamental Bragg reflection peaks in a given sample, using the relation[10] $S^2 = (I_S/I_F)_{sample} \times (I_F/I_S)_{s=1}$, where $(I_F/I_S)_{s=1}$ is the corresponding intensity ratio for the fully ordered LaAg compound. The ratio $(I_F/I_S)_{S=1}$ is calculated from the expression

$$(I_F/I_S)_{S=1} = \frac{\{X_{La}f_{La}(\theta) + X_{Ag}f_{Ag}(\theta)\}_F^2 \{L_P(\theta)\}_F \{\exp(-2M(\theta))\}_F}{\{f_{La}(\theta) - f_{Ag}(\theta)\}_S^2 \{L_P(\theta)\}_S \{\exp(-2M(\theta))\}_S} \qquad (3)$$

where $X_{La}$ and $X_{Ag}$ are the concentrations of La and Ag atoms in atomic fractions, $f_{La}$ and $f_{Ag}$ are the structure factors for the La and Ag sublattices, $L_P = 1/2\sin^2\theta\cos\theta$ is the Lorentz polarization factor, *2θ* is the Bragg angle and $\exp(-2M)$ is the Debye-Waller factor with $M(\theta) = (3h^2/m_a k_B \Theta_D)\{(\phi(x)/x) + (1/4)\}(\sin\theta/\lambda)^2$, Debye integral $\phi(x) = (1/x)\int_0^x \xi\, d\xi/(e^\xi - 1)$, Debye temperature $\Theta_D$, $x = \Theta_D/T$ and mass of the atom $m_a$. In equation (3), the subscripts F and S refer to the fundamental and superstructure reflections, respectively. The theoretical estimate for the integrated intensity ratios $[I_F^{(200)}/I_S^{(100)}]_{S=1}$ and $[I_F^{(222)}/I_S^{(111)}]_{S=1}$ have been obtained from the standard neutron diffraction pattern for fully ordered LaAg, calculated by the Rietveld method. The long-range order parameter has then been calculated from the relations $S^2 = [I_S^{(100)}/I_F^{(200)}]_{meas} \times [I_F^{(200)}/I_S^{(100)}]_{S=1}$ and $S^2 = [I_S^{(111)}/I_F^{(222)}]_{meas} \times [I_F^{(222)}/I_S^{(111)}]_{S=1}$, using the measured (denoted by subscript 'meas') integrated intensity ratios of the (100), (200) and (111), (222) reflections for a given sample. The values of S, so obtained, are listed in Table II and plotted against the Mn concentration in figure 11.

In the completely ordered LaAg compound, La (Ag) atoms are found only at the body-centred (corner) sites so that the LaAg lattice is composed of the La and Ag sublattices. In general, the long-range (atomic) order parameter is defined as $S = (r-w)/(r+w)$, where *r* and *w* represent the number of right atoms (e.g. La atoms on La sites or Ag atoms on Ag sites) and wrong atoms (e.g. La and/or Mn atoms on Ag sites or Ag and/or Mn atoms on La sites). Thus, S attains its maximum value S = 1 only for the completely ordered compound LaAg when $w = 0$ and is bound to be less than unity when the Mn concentration is non-zero (since Mn atoms are wrong atoms irrespective of whether they occupy La or Ag sites). Thus regardless of sample thermal history and the site preference of Mn, S is expected to decrease monotonously with increasing Mn concentration (*x*), in agreement with the present observation (Fig. 11), barring the sudden dip at $x = c = 0.05$. The most striking feature of the above definition of the long-range order parameter is that it permits an accurate determination of the site occupation of the La and Ag sublattices. To elucidate this point, we take the example of the alloy with c = 0.1 for which the observed value of S is $S_{obs}$ = 0.90(1) (Table II). For each sublattice (La or Ag), $r + w = 100$ and hence the value $S_{cal} = (r-w)/(r+w)$ yields $r - w = 90$. It immediately follows that there are 95 *right* atoms and 5 *wrong* atoms on a given sublattice (La or Ag).

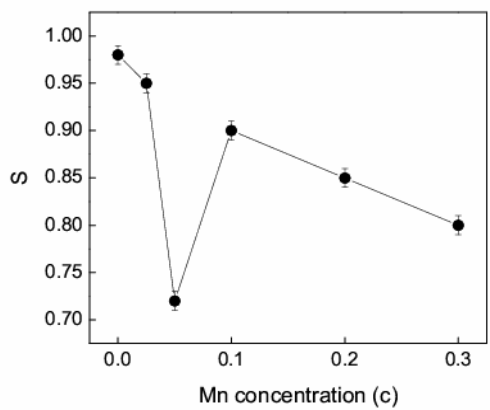

FIG.11. Long range order parameter S as a function of actual Mn concentration x.

Thus, the maximum number of Mn atoms that each of the sublattices can accommodate is 5. If 10 Mn atoms are equally distributed among La and Ag sublattices, the expected composition $La_{0.95}Ag_{0.95}Mn_{0.10}$ of the alloy is *at variance* with the observed composition $La_{0.95}Ag_{1.0}Mn_{0.05}$ (Table I). Similarly, other possibilities can be ruled out because the predicted composition does not agree with the observed one. Of all the possible ways (Table II lists only a small sub-set of such possibilities) of arranging La, Ag and Mn atoms on the La and Ag sublattices that are consistent with the observed value of S, the only distribution of La, Ag and Mn atoms on La (Ag) sublattice that yields the observed composition $La_{0.95}Ag_{1.0}Mn_{0.05}$ (Table II) is 95, 5, 0 (0, 95, 5). This unique choice of atomic distribution implies that 5 Mn atoms replace 5 Ag atoms on the Ag sublattice and the 5 Ag atoms, so displaced, land up on the La sublattice. Consequently, half of the Mn concentration does not form a part of the CsCl lattice of LaAg and segregates out.

The long-range order parameter, S, as a function of Mn concentration (Fig. 11) goes through a steep dip (S = 0.72) at $x = c = 0.05$. It follows from this observation that the alloy with $c$ = 0.05 is more site-disordered than the other compositions. The above calculation when repeated with the observed value of S, $S_{obs}$ = 0.72(1), yields 86 right atoms and 14 wrong atoms on each sublattice (La or Ag). Considering that the observed composition of this alloy is $La_{0.95}Ag_{1.0}Mn_{0.05}$, the total number of Mn atoms on both the sublattices cannot exceed 5. Thus out of the six possible ways Mn atoms can occupy the La and Ag sublattices, only two distributions of La, Ag and Mn atoms, i.e., 86, 14, 0 (9, 86, 5) and 86, 9, 5 (14, 86, 0), on La (Ag) sublattice correspond to the compositions $La_{0.95}Ag_{1.0}Mn_{0.05}$ (where Mn replaces Ag on the Ag sites) and $La_{1.0}Ag_{0.95}Mn_{0.05}$ (where Mn replaces La on the La sites) observed in the white and dark regions in the scanning electron micrograph (Fig. 1(c)). This inference permits us to conclude that, in this alloy, all the Mn atoms either occupy La sites or Ag sites. Table 2 displays the observed and calculated values of the order parameter S, possible La and Ag site occupancies (by La, Ag and Mn atoms), as well as the predicted and observed compositions of the alloys. The following conclusions regarding the Mn site preference can be drawn from the entries in Table II. (I) The site preference of Mn depends strongly on the Mn composition; Mn has exclusive La (Ag) site preference in the alloy (alloys) with $x = c = 0.025$ ($c \geq 0.1$) whereas it has essentially no site preference in the alloy with $c = 0.05$ where all the Mn atoms are either found on La sites or on Ag sites. That the exclusive La site preference of Mn in the alloy with $x = c = 0.025$ is consistent with the observation that the lattice parameter has a much higher value for this composition than in the neighbouring compositions $c = 0$ and $c = 0.05$ becomes clear when cognizance is taken of the following fact. Due to a strong repulsive potential between the La and Mn ions (as is indicated by the immiscibility[11] of Mn in La even in the liquid state), the CsCl lattice of LaAg is expected to dilate when Mn atoms occupy La sites. By contrast, when Mn substitutes Ag on the Ag sites, as is the case for the alloys with $x \geq 0.05$ (or $c \geq 0.1$), the lattice parameter follows a linear variation with $x$ (Vegard's law) because Ag and Mn form homogeneous solid solutions[11] up to 30 at. % Mn in Ag. (II) Though the observed values of the long-range order parameter S, $S_{obs}$, set the upper bounds on the concentration of Mn in the alloys with c = 0.025, 0.05, 0.1, 0.2 and 0.3 as 0.025, 0.05, 0.1, 0.16 and 0.2, respectively, the actual Mn concentration in the *last three alloys* is exactly half of these maximum values. This is a consequence of the fact that Mn has a strong tendency to segregate out of the CsCl lattice of the host, LaAg.

**Table II**. Long-range order parameter S, population of La and Ag sublattices, predicted and observed composition.

| Sample | $S_{obs}$ | $S_{cal}$ | La sublattice | | | Ag sublattice | | | Composition | |
|---|---|---|---|---|---|---|---|---|---|---|
| | | | La | Ag | Mn | La | Ag | Mn | Predicted | Observed |
| c = 0.0 | 0.98(2) | 0.98 | 99.0 | 1.0 | 0.0 | 1.0 | 99.0 | 0.0 | LaAg | $La_{1.00(1)}Ag_{1.00(1)}$ |
| c = 0.025 | 0.95(1) | 0.95 | 97.5 | 0.0 | 2.5 | 2.5 | 97.5 | 0.0 | $La_{1.000}Ag_{0.975}Mn_{0.025}$ | $La_{1.000(5)}Ag_{0.975(5)}Mn_{0.025(1)}$ |
| | | | 97.5 | 2.5 | 0.0 | 0.0 | 97.5 | 2.5 | $La_{0.975}Ag_{1.0}Mn_{0.025}$ | |
| | | | 97.5 | 0.5 | 2.0 | 2.0 | 97.5 | 0.5 | $La_{0.995}Ag_{0.98}Mn_{0.025}$ | |
| | | | 97.5 | 2.0 | 0.5 | 0.5 | 97.5 | 2.0 | $La_{0.980}Ag_{0.995}Mn_{0.025}$ | |
| c = 0.05 | 0.72(1) | 0.72 | 86.0 | 14.0 | 0.0 | 9.0 | 86.0 | 5.0 | $La_{0.95}Ag_{1.0}Mn_{0.05}$ | $La_{0.950(1)}Ag_{1.000(1)}Mn_{0.050(2)}$ |
| | | | 86.0 | 9.0 | 5.0 | 14.0 | 86.0 | 0.0 | $La_{1.00}Ag_{0.95}Mn_{0.05}$ | $La_{1.00(1)}Ag_{0.950(1)}Mn_{0.050(2)}$ |
| | | | 86.0 | 13.0 | 1.0 | 13.0 | 86.0 | 4.0 | $La_{0.96}Ag_{0.99}Mn_{0.05}$ | |
| | | | 86.0 | 10.0 | 4.0 | 13.0 | 86.0 | 1.0 | $La_{0.99}Ag_{0.96}Mn_{0.05}$ | |
| c = 0.1 | 0.90(1) | 0.90 | 95.0 | 0.0 | 5.0 | 0.0 | 95.0 | 5.0 | $La_{0.95}Ag_{0.95}Mn_{0.10}$ | |
| | | | 95.0 | 5.0 | 0.0 | 0.0 | 95.0 | 5.0 | $La_{0.95}Ag_{1.0}Mn_{0.05}$ | $La_{0.95(1)}Ag_{1.00(1)}Mn_{0.050(2)}$ |
| | | | 95.0 | 0.0 | 5.0 | 5.0 | 95.0 | 0.0 | $La_{1.00}Ag_{0.95}Mn_{0.05}$ | |
| | | | 95.0 | 4.0 | 1.0 | 1.0 | 95.0 | 4.0 | $La_{0.96}Ag_{0.99}Mn_{0.05}$ | |
| | | | 95.0 | 1.0 | 4.0 | 4.0 | 95.0 | 1.0 | $La_{0.99}Ag_{0.96}Mn_{0.05}$ | |
| c = 0.2 | 0.85(1) | 0.84 | 92.0 | 0.0 | 8.0 | 0.0 | 92.0 | 8.0 | $La_{0.92}Ag_{0.92}Mn_{0.16}$ | |
| | | | 92.0 | 8.0 | 0.0 | 0.0 | 92.0 | 8.0 | $La_{0.92}Ag_{1.0}Mn_{0.08}$ | $La_{0.92(1)}Ag_{1.00(1)}Mn_{0.080(2)}$ |
| | | | 92.0 | 0.0 | 0.0 | 0.0 | 92.0 | 8.0 | $La_{1.00}Ag_{0.92}Mn_{0.08}$ | |
| | | | 92.0 | 7.0 | 1.0 | 1.0 | 92.0 | 7.0 | $La_{0.93}Ag_{0.99}Mn_{0.08}$ | |
| | | | 92.0 | 1.0 | 7.0 | 7.0 | 92.0 | 1.0 | $La_{0.99}Ag_{0.93}Mn_{0.08}$ | |
| = 0.3 | 0.80(1) | 0.8 | 90.0 | 0.0 | 10.0 | 0.0 | 90.0 | 10.0 | $La_{0.90}Ag_{0.90}Mn_{0.2}$ | |
| | | | 90.0 | 10.0 | 0.0 | 0.0 | 90.0 | 10.0 | $La_{0.90}Ag_{1.0}Mn_{0.1}$ | $La_{0.900(5)}Ag_{1.000(5)}Mn_{0.100(1)}$ |
| | | | 90.0 | 0.0 | 10.0 | 10.0 | 90.0 | 0.0 | $La_{1.00}Ag_{0.90}Mn_{0.1}$ | |
| | | | 90.0 | 9.0 | 1.0 | 1.0 | 90.0 | 9.0 | $La_{0.91}Ag_{0.99}Mn_{0.1}$ | |
| | | | 90.0 | 1.0 | 9.0 | 9.0 | 90.0 | 1.0 | $La_{0.99}Ag_{0.91}Mn_{0.1}$ | |

### D. Superconducting-normal phase diagram

Previous ac susceptibility, electrical resistivity, $\rho(T)$, and the specific heat, $C_P(T)$, measurements[1] indicated that the superconducting transition temperature, $T_C$, lies below 1.7 K (the lowest temperature reached in these experiments) for the alloys with $c = x = 0$ and $c = x = 0.025$. This observation prompted us to extend the above measurements, particularly for these two concentrations, to 0.35 K. Figure 12 displays the *revised* superconducting-normal phase diagram, $T_C(x)$, based on the new $\rho(T)$ and $C_P(T)$ data taken on LaAg and LaAg$_{0.975}$Mn$_{0.025}$ samples at temperatures down to 0.35 K. The observed variation of $T_C$ with $x$ (Fig. 12), when viewed against the backdrop of the EDAX, XRD and ND results, leads us to the following obvious conclusions. The conventional phonon-mediated superconductivity prevalent in the parent LaAg compound at $T \leq T_C \cong 1K$ is completely suppressed in the alloy with $c = x = 0.025$ when Mn substitutes La at the La sub-lattice sites in the LaAg host. By contrast, the substitution of Ag by Mn on the Ag sub-lattice sites in LaAg, as is the case in the alloys with $x \geq 0.05$, gives rise to unconventional superconductivity, which cannot be explained[1] in terms of the BCS phonon-mediated pairing mechanism. Thus, there is a direct correlation between the site preference of Mn in LaAg and the nature of superconductivity.

That there is a threshold Mn concentration $x = 0.05$ at which the unconventional superconductivity first appears in the $LaAg_{1-x}Mn_x$ alloys is evident from the following observations. The 'as-cast' $c = x = 0.05$ sample, which was superconducting at temperatures[1] $T \leq T_c \approx 5K$, *ceases* to be a superconductor after it is annealed and exhibits a behavior exactly identical[1] to that of the 'as-prepared' $c = x = 0.025$ sample. The EDAX result that the majority phase in 'annealed' c = 0.05 sample has the same Mn content as in the 'as-prepared' c = 0.025 leads to the obvious conclusion that the unconventional superconductivity is not observed when the Mn concentration falls to 2.5 at. %. Moreover, the finding that the 'as-cast' $c = x = 0.05$ and $c = 0.1$ alloys have exactly the *same*[1] superconducting transition temperature, $T_C$ = 5 K, is not surprising considering the EDAX analysis result that the majority phase in c = 0.1 alloy has essentially the same composition as that of the $c = x = 0.05$ alloy. The above observations thus strongly suggest that the 3d electrons of Mn could be responsible for the unconventional superconductivity observed in the $LaAg_{1-x}Mn_x$ alloys with $0.05 \leq x \leq 0.1$.

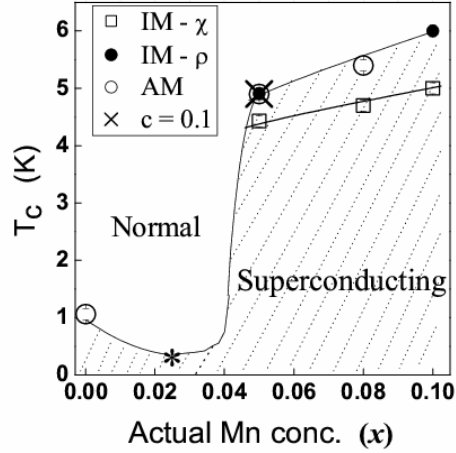

FIG.12. Superconducting-normal phase diagram. The value of $T_C$ in the alloy with $c = x = 0.025$ corresponds to the temperature which marks the *onset* of superconductivity.

## IV. SUMMARY AND CONCLUSION

The crystallographic structure and composition of the ternary alloy $LaAg_{1-c}Mn_c$ with c = 0.0, 0.025, 0.05, 0.1, 0.1 and 0.3, has been investigated by x-ray diffraction (XRD), neutron diffraction (ND), scanning electron microscopy (SEM) and energy dispersive absorption of x-rays (EDAX). The EDAX analysis reveals that the alloys with the nominal Mn concentration $c = 0$, 0.025 and 0.05 are *essentially single phase* alloys with the actual Mn concentration, *x*, *same as* the nominal one, i.e., $c = x$, whereas in the alloys with $c = 0.1$, 0.2 and 0.3, the actual Mn concentration of the majority phase (crystalline grains) is $x = 0.050(1)$, 0.080(1) and 0.100(1), respectively. The room temperature XRD patterns and ND data taken over a wide temperature range (1.8 K to 50 K) reveal that the CsCl structure of the parent compound LaAg is retained in all the Mn containing alloys over the whole temperature range and all the minority phases put together are less than 1 volume percent of the majority phase in the alloys with $0.05 \leq x \leq 0.1$ (or $0.1 \leq c \leq 0.3$).

Even in the alloy with the highest Mn content, ND patterns, in the temperature range $1.8\,K \leq T \leq 50\,K$, do not exhibit any superstructure peak(s) at low momentum transfer values that normally reflect the existence of long-range antiferromagnetic order. At a fixed temperature, the lattice parameter '$a$' as a function of $x$ goes through a peak at $x = 0.025$ and this peak is followed by *a linear variation* of $a$ with $x$ for $x \geq 0.05$. Consistent with this composition dependence of the lattice parameter, the observed values of the long-range atomic order parameter for different compositions assert that Mn has *exclusive La (Ag) site preference* in the alloy (alloys) with $x = c = 0.025$ ($x \geq 0.05$) whereas in the alloy with $c = 0.05$, Mn has essentially no site preference in that all the Mn atoms either occupy the La sites or the Ag sites. Unconventional superconductivity in the LaAg$_{1-x}$Mn$_x$ alloy system appears only when Mn substitutes Ag at the Ag sites, as is the case in the alloys with $x \geq 0.05$. At low Mn concentrations $x \approx 0.025$, substitution of La by Mn at the La sub-lattice sites in LaAg host essentially destroys the conventional phonon-mediated superconductivity prevalent in the parent LaAg compound.

## ACKNOWLEDGEMENT

The authors thank A. Señas (L. Sanchez Aramburu) for rendering assistance in the sample preparation (in taking scanning electron micrographs and for EDAX analysis). We thank Institute Laue-Langevin for the allocation of time on the D20 diffractometer. S. Kumar thanks Council for Scientific and Industrial Research, India, for granting the research fellowship. This work was supported by MEC research project no. MAT 2005-6806-C04-02 and SAB2001-0086, Spain.